\documentstyle[11pt]{article}

\newcommand{\sect}[1]{ \section{#1} \setcounter{equation}{0} }
\newcommand{\partialslash}{\partial \! \! \! /}

\newcommand{\xslash}{x \! \! \! /}

\newcommand{\half}{\mbox{\small{$\frac{1}{2}$}}}

\newcommand{\la}{\langle}
\newcommand{\ra}{\rangle}
\newcommand{\Nf}{N_{\!f}}
\begin{document}
\title{Algorithm for computing the $\beta$-function of quantum electrodynamics
in the large $\Nf$ expansion.}
\author{J.A. Gracey, \\ Department of Applied Mathematics and Theoretical
Physics, \\ University of Liverpool, \\ P.O. Box 147, \\ Liverpool, \\
L69 3BX, \\ United Kingdom.}
\date{}
\maketitle
\vspace{5cm}
\noindent
{\bf Abstract.} By considering corrections to the asymptotic scaling functions
of the photon and electron in quantum electrodynamics with $\Nf$ flavours, we
solve the skeleton Dyson equations at $O(1/\Nf)$ in the large $\Nf$ expansion
at the $d$-dimensional critical point of the theory and deduce the critical
exponent $\beta^\prime(g_c)$, in arbitrary dimensions, and subsequently
present a method for computing higher order corrections to $\beta(g)$.

\vspace{-16cm}
\hspace{10cm}
{\bf LTH-294}
\newpage
\sect{Introduction.}
Recently, various authors have provided an insight into the four loop
structure of quantum electrodynamics, (QED), by calculating the
$\beta$-function to this order, \cite{1}. This renormalization group function,
which describes how the renormalized coupling constant depends on the
renormalization scale, is important in relating theoretical predictions of the
theory with experiment. From a technical point of view, the computation of the
functions which arise in the renormalization group equation, and also include,
for example, the anomalous dimension of the fundamental fields, become quite
intricate after several orders, \cite{1}. Indeed the latest calculations could
only be completed in a reasonable time with the use of computer algebra
packages, \cite{2}. To push the accuracy of current perturbative calculations
beyond the present four loop status, an important aid is the provision of
independent (analytic) analysis. In general, this is not usually possible for
all renormalizable quantum field theories. For a subclass of theories which
possess an internal symmetry, however, it is possible to carry out different
perturbative analysis, which can then be related to the conventional
coupling constant perturbation series. For instance, if a theory possesses an
$O(N)$ or $U(N)$ internal symmetry, one can let the parameter $N$ become
large, whence $1/N$ becomes small, and therefore reorder the perturbation
series to analyse all $O(1/N)$ graphs initially, before proceeding to the
subsequent order. More specifically, this large $N$ expansion corresponds to
reordering perturbation theory, such that one sums chains of bubble graphs
first. As we will be discussing QED in detail here, we note that with $\Nf$
flavours of electrons this theory indeed possesses a large $\Nf$ expansion.

Earlier studies of QED in a large $\Nf$ expansion were carried out in
\cite{3,4}, where the authors performed the explicit bubble sum for the photon
and then used it to renormalize the one loop graphs in minimal subtraction,
(MS), using dimensional regularization in $d$ $=$ $4$ $-$ $2\epsilon$
dimensions. In this way, the $\beta$-function as well as the renormalization
group function corresponding to the electron mass were deduced at $O(1/\Nf)$.
Indeed this independent result was used in \cite{1} to check the $4$-loop term
where possible. More recently, the electron anomalous dimension was computed in
the Landau gauge to $O(1/\Nf)$ in \cite{5} and later to $O(1/\Nf^2)$ in
\cite{6}. These calculations were carried out without having to explicitly sum
any bubble graphs and perform the usual renormalization. Instead the model was
analysed at the $d$-dimensional critical point where the fields obey asymptotic
scaling and are conformal, and the theory is finite. Consequently, taking a
general power law structure for the fields in the neighbourhood of criticality,
where the (critical) exponent depends only on $\Nf$ and $d$, and are related to
the renormalization group function corresponding to the electron anomalous
dimension, $\gamma(g)$, it was possible to deduce the critical exponent
algebraically by solving the truncated skeleton Dyson equation with dressed
propagators. Indeed this approach was based on the pioneering work of
\cite{7,8} which dealt with the $O(N)$ bosonic $\sigma$ model and extensions to
other models, \cite{9,10}. Knowledge of these exponents in arbitrary dimensions
gives an indirect way of determining the renormalization group functions. Since
the exponent can be expanded in powers of $\epsilon$ and the critical coupling
$g_c$ $\sim$ $3 \epsilon/\Nf$, taking a general form for say $\gamma(g)$, its
undetermined $n$th order coefficient can be deduced from the coefficient of
$\epsilon^n$ in the expansion of the critical exponent.

In this paper, we will extend the formalism of \cite{6,8} to compute the
$\beta$-function critical exponent for QED at $O(1/\Nf)$, which is important
for several reasons. First, it will, of course, provide an independent check on
the result of \cite{3} and also ensure that the method we will use is correct.
As far as we are aware, the proper treatment of the corrections to the
asymptotic scaling functions of a $U(1)$ gauge field has not been given
previously. Having demonstrated its usefulness here, it will open up the
possibility of carrying out further calculations in other models, such as the
bosonic $CP(N)$ $\sigma$ model, which possess a $U(1)$ gauge field. Second,
and more importantly, the formalism we introduce, will provide the foundation
for probing the $\beta$-function to $O(1/\Nf^2)$ and we will outline several
technical points in formulating the algorithm to carry this out. Indeed, we
suspect the method of \cite{8}, which has proved successful in providing the
only $O(1/N^2)$ information of critical exponents in other models, will provide
a quicker evaluation from a technical point of view than \cite{3}, since, for
example, one does not have to compute bubble sum corrections with massive
electrons.

The paper is organised as follows. In sect. 2, we introduce the formalism
we will use, discussing in detail the derivation of the asymptotic scaling
form of the propagators and $2$-point functions of the various fields. These
are used in sect. 3, to represent the Dyson equations at criticality by
algebraic equations, which can be solved systematically at each order in
$1/\Nf$. After pointing out the need to include a higher order two loop
graph, which is a novel feature for fermionic theories compared with the
treatment of bosonic theories of \cite{7,8}, we proceed to compute it in sect.
4, where we give an extensive discussion about how to calculate certain
massless two loop Feynman diagrams, using a set of recursion relations. The
full calculation is finally assembled in sect. 5, where we outline certain
technical points concerning the $O(1/\Nf^2)$ evaluation of the
$\beta$-function. An appendix contains a library of the basic results required
at certain stages.

\sect{Preliminaries.}
We begin by introducing the notation and basic formalism for analysing QED
with $\Nf$ flavours of electrons at its critical point, defined as the
non-trivial $d$-dimensional zero of $\beta(g)$. First, we recall the
(massless) lagrangian is
\begin{equation}
L ~=~ i \bar{\psi}^i \partialslash \psi^i + A_\mu \bar{\psi}^i \gamma^\mu
\psi^i - \frac{(F_{\mu\nu})^2}{4e^2} - \frac{(\partial_\mu A^\mu)^2}
{2be^2}
\end{equation}
where $1$ $\leq$ $i$ $\leq$ $\Nf$, $A_\mu$ is the $U(1)$ gauge field with
$F_{\mu\nu}$ $=$ $\partial_\mu A_\nu$ $-$ $\partial_\nu A_\mu$ and $b$ is the
covariant gauge parameter. We have rescaled the electron charge, $e$, into the
quadratic term of the gauge field to ensure there is a unit coupling at the
$3$-vertex, in anticipation of applying the methods of \cite{7,8}. However, we
will use $g$ $=$ $(e/2\pi)^2$ as our coupling constant here, following the
conventions of \cite{11}. Next, we write down the asymptotic scaling form of
the propagators of the fields in the critical region, where the theory is
massless and has a conformal symmetry. In earlier work, we considered only the
leading order form of these functions which depend on the full dimensions of
the fields. Here, we go beyond the leading order of \cite{6} and consider
corrections, where, as was discussed in other models, \cite{8}, the critical
exponent involved the slope of the $\beta$-function at criticality. Thus, we
take the following coordinate space forms for the propagators as
\begin{eqnarray}
\psi(x) & \sim & \frac{A\xslash}{(x^2)^\alpha} [ 1 + (x^2)^\lambda A^\prime]
\\
A_{\mu \nu} (x) & \sim & \frac{B}{(x^2)^\beta} \left[ \eta_{\mu \nu}
+ \frac{2\beta}{(2\mu-2\beta-1)} \frac{x_\mu x_\nu}{x^2} \right. \nonumber \\
&+& \left. (x^2)^\lambda B^\prime \left( \eta_{\mu\nu} +
\frac{2(\beta-\lambda)}
{(2\mu-2\beta+2\lambda-1)} \frac{x_\mu x_\nu}{x^2} \right) \right]
\end{eqnarray}
The quantities $(A,B)$ and $(A^\prime, B^\prime)$ are amplitudes which are
independent of $x$, whilst $\alpha$ and $\beta$ are the dimensions or
critical exponents of the electron, $\psi^i$, and $A_\mu$ respectively and
$2\lambda$ $=$ $- \, \beta^\prime(g_c)$, which is a gauge independent
quantity.

Several comments concerning (2.3) are in order. First, since we are dealing
with a gauge field, we have to choose a gauge. As we are carrying out a
large $\Nf$ expansion, this constrains us to use the Landau gauge, because
from a perturbative point of view renormalization effects are such that one
remains in this gauge, which is not the case for other covariant gauges.
Since the large $\Nf$ expansion is a reordering of perturbation theory such
that one sums chains of bubble graphs first, choosing any other covariant gauge
means one cannot compare results from the large $\Nf$ expansion with
perturbation theory due to the mixing of gauges which would occur, \cite{3,4}.
Secondly, with this choice, we note that one can recover the normal Landau
gauge structure of $A_{\mu\nu}$ by transforming (2.3) to momentum space via the
Fourier transform, \cite{7},
\begin{equation}
\frac{1}{(x^2)^\alpha} ~=~ \frac{a(\alpha)}{2^{2\alpha}\pi^\mu} \int_k
\frac{e^{ikx}}{(k^2)^{\mu-\alpha}}
\end{equation}
valid for all $\alpha$, and its derivatives, where $a(\alpha)$ $=$
$\Gamma(\mu-\alpha)/\Gamma(\alpha)$, whence (2.3) takes the form
\begin{equation}
\frac{\tilde{B}}{(k^2)^{\mu-\alpha}} \left[ \eta_{\mu\nu} - \frac{k_\mu k_\nu}
{k^2} \right] \left[ 1 + \frac{\tilde{B}^\prime}{(k^2)^\lambda} \right]
\end{equation}
This also illustrates our next point concerning the form of the corrections we
take for the asymptotic scaling form of the $A_\mu$ propagator and why the
form of (2.3) might have appeared contrived. In fact, we have taken the
simplest correction for $A_{\mu\nu}$ in momentum space having been motivated
partly by the way one deals with bosonic and fermionic fields.

{}From a dimensional analysis of the action, we define, \cite{5},
\begin{equation}
\alpha ~=~ \mu + \half \eta ~~~,~~~ \beta ~=~ 1 - \eta - \chi
\end{equation}
where $\mu$ $=$ $d/2$, which introduces the electron anomalous dimension,
$\eta$, and the anomalous dimension of the $3$-vertex of (2.1), $\chi$. Both
are $O(1/\Nf)$ within the large $\Nf$ expansion and have been calculated to
$O(1/\Nf^2)$ and $O(1/\Nf)$, respectively, in the Landau gauge in
\cite{6,7,13},
by considering only the leading order terms of (2.2) and (2.3), ie
\begin{equation}
\eta_1 ~=~ - \, \chi_1 ~=~ - \, \frac{(2\mu-1)(2-\mu)\Gamma(2\mu)}
{4\Gamma^2(\mu) \Gamma(\mu+1) \Gamma(2-\mu)}
\end{equation}
which is all we require for this paper, where $\eta$ $=$ $\sum_{i=1}^\infty
\eta_i/\Nf^i$.

A final piece of formalism required for the Dyson equation is the scaling
forms of the $2$-point functions, denoted by $\psi^{-1}(x)$ and
$A_{\mu\nu}^{-1}(x)$. The formula for a fermion field was introduced in
\cite{10} as
\begin{equation}
\psi^{-1}(x) ~ \sim ~ \frac{r(\alpha-1)\xslash}{A(x^2)^{2\mu-\alpha+1}}
[ 1 - A^\prime s(\alpha-1, \lambda) (x^2)^\lambda]
\end{equation}
where
\begin{equation}
r(\alpha) ~=~ \frac{\alpha a(\alpha-\mu)}{(\mu-\alpha)a(\alpha) \pi^{2\mu}}
{}~~,~~~ s(\alpha, \lambda) ~=~ \frac{\alpha(\alpha-\mu) q(\alpha,\lambda)}
{(\alpha-\mu+\lambda)(\alpha-\lambda)}
\end{equation}
with $q(\alpha,\lambda)$ $=$ $a(\alpha-\mu+\lambda)a(\alpha-\lambda)/a(\alpha
-\mu)a(\alpha)$, and is derived as follows. One first maps (2.2) to momentum
space and inverts the scaling function on this space via $G^{-1}G$ $=$ $1$, to
give the momentum space form of the $2$-point function, before converting back
to coordinate space, \cite{7}. In the case of the gauge field one restricts the
inversion to the transverse subspace of momentum space since this is the only
physically relevant piece, \cite{13,14}, and therefore
\begin{eqnarray}
A^{-1}_{\mu\nu}(x) & \sim & \frac{m(\beta)}{B(x^2)^{2\mu-\beta}} \left[
\eta_{\mu\nu} + \frac{2(2\mu-\beta)}{(2\beta-2\mu-1)} \frac{x_\mu x_\nu}{x^2}
\right. \nonumber \\
& - & \left. (x^2)^\lambda B^\prime n(\beta) \left( \eta_{\mu\nu}
+ \frac{2(2\mu-\beta-\lambda)}{(2\beta+2\lambda-2\mu-1)} \frac{x_\mu x_\nu}
{x^2} \right) \right]
\end{eqnarray}
where
\begin{eqnarray}
m(\beta) & = & \frac{[4(\mu-\beta)^2-1]a(\beta-\mu)}{4\pi^{2\mu}(\mu-\beta)^2
a(\beta)} \nonumber \\
n(\beta) & = & \frac{(\mu-\beta+\lambda)(2\mu-2\beta-1)(2\mu
-2\beta-2\lambda+1)}{(\mu-\beta-\lambda)(2\mu-2\beta+1)(2\mu-2\beta
+2\lambda-1)} q(\beta,\lambda) ~~~
\end{eqnarray}
Finally, for completeness we note that the $4$-loop $\beta$-function for (2.1)
is in $\overline{\mbox{MS}}$, \cite{1,15,16,17}
\begin{eqnarray}
\beta(g) &=& (d-4)g + \frac{2\Nf}{3} g^2 + \frac{\Nf}{2} g^3
- \frac{\Nf(22\Nf+9)}{144} g^4 \nonumber \\
&-& \frac{\Nf}{64} \left[ \frac{616\Nf^2}{243} + \left( \frac{416 \zeta(3)}{9}
- \frac{380}{27} \right)\Nf + 23 \right] g^5 + O(g^6) ~~~~
\end{eqnarray}
Thus, the location of the non-trivial $d$-dimensional critical point, $g_c$, is
\begin{equation}
g_c ~=~ \frac{3\epsilon}{\Nf} - \frac{27 \epsilon^2}{4\Nf^2}
+ \frac{99\epsilon^3}{16\Nf^2} + \frac{77\epsilon^4}{16\Nf^2} + O(\epsilon^5)
\end{equation}

\sect{Consistency equation for $\lambda$.}

In this section, we discuss the derivation from the Dyson equation of the
consistency equation whose solution will yield the $O(1/\Nf)$ correction
to $\lambda$ and will therefore become the master equation for setting up the
algorithm for $O(1/\Nf^2)$ corrections. Moreover, as will be seen, the
derivation is not as straightforward as in other models.

First, the skeleton Dyson equations with dressed propagators we initially
consider are illustrated in fig. 1, which have been truncated to the order in
large $\Nf$ we presently consider. Since we have introduced our formalism of
the previous section, in coordinate space, one can therefore represent fig. 1
at the critical point, (2.13), immediately, by replacing the lines of the one
loop graphs by (2.2) and (2.3), without the need to compute any one loop
integrals. Thus,
\begin{eqnarray}
0 &=& r(\alpha-1) [ 1 - s(\alpha-1,\lambda) (x^2)^\lambda A^\prime ] \\
&+& 2(2\mu-1)z \left[ \frac{(\beta-\mu+1)}{(2\mu-2\beta-1)}
[1 + (x^2)^\lambda A^\prime] + \frac{(x^2)^\lambda B^\prime (\beta-\mu
-\lambda+1)}{(2\mu-2\beta+2\lambda-1)} \right] \nonumber \\
0 &=& A^{-1}_{\mu\nu}(x) - \frac{4z\Nf}{(x^2)^{2\alpha-1}} \left( \eta_{\mu\nu}
- \frac{2x_\mu x_\nu}{x^2} \right) [ 1 + 2 A^\prime (x^2)^\lambda ]
\end{eqnarray}
where $z$ $=$ $A^2B$. As we noted earlier, only the transverse part of the
Dyson equation for $A_\mu$ in momentum space is physically relevant. So,
following \cite{13,14}, we isolate this piece of (3.2) by first transforming
(3.2) to momentum space, noting that we have not cancelled the powers of $x^2$,
projecting out the transverse component before applying the inverse mapping.
Equivalently, one can make the replacement
\begin{equation}
\frac{x_\mu x_\nu}{(x^2)^\alpha} ~ \longrightarrow ~ \frac{\eta_{\mu\nu}}
{2(\alpha-1) (x^2)^{\alpha-1}}
\end{equation}
valid for all $\alpha$, in (3.2) before discarding the isotropic tensor and
then cancelling powers of $x^2$. Thus
\begin{eqnarray}
0 &=& m(\beta) \left[ \frac{(\beta-\mu)}{(2\beta-2\mu-1)} - \frac{B^\prime
(x^2)^\lambda n(\beta) (\beta+\lambda-\mu)}{(2\beta-2\mu+2\lambda-1)}
\right] \nonumber \\
&-& 4z\Nf \left[ \frac{(\alpha-1)}{(2\alpha-1)} + \frac{A^\prime (x^2)^\lambda
(2\alpha-\lambda-2)}{(2\alpha-\lambda-1)} \right]
\end{eqnarray}
Next, since both (3.1) and (3.4) contain terms with $(x^2)^\lambda$ and either
$A^\prime$ or $B^\prime$, we treat these portions as independent and decouple
each equation into two. The terms corresponding to the leading terms of (2.2)
and (2.3) give two consistency equations whose solution yield $\eta_1$. The
remaining two equations correspond, naively, to the consistency equation whose
solution ought to give $\lambda_1$, by setting the determinant of the $2$
$\times$ $2$ matrix formed by the two remaining equations, to zero. These are
\begin{eqnarray}
0 &=& [1 + s(\alpha-1,\lambda)]A^\prime ~+~ \frac{(2\mu-2\beta-1)(\beta-\mu
-\lambda+1) B^\prime}{(2\mu-2\beta+2\lambda-1)(\beta-\mu+1)} \\
0 &=& \frac{(2\alpha-1)(2\alpha-\lambda-2) A^\prime}{(2\alpha-\lambda-1)
(\alpha-1)} ~+~ \frac{(\mu-\beta-\lambda)(2\mu-2\beta+1)n(\beta)
B^\prime}{(\mu-\beta)(2\mu-2\beta-2\lambda+1)} ~~~~~
\end{eqnarray}

Until this point we have directly followed the formulation used in the bosonic
and supersymmetric $O(N)$ $\sigma$ models to determine $\lambda$, \cite{8,10}.
However, unlike these models there is a fundamental difference in the QED
formulation. With $\lambda$ $=$ $\mu$ $-$ $2$ $+$ $\sum_{i=1}^\infty
\lambda_i/\Nf^i$,
\begin{equation}
s(\alpha-1, \lambda) ~=~ \frac{4(\mu-2)\Nf}{\mu \eta_1} ~+~ O(1)
\end{equation}
which is $O(\Nf)$. However, from (3.6) the coefficient of $B^\prime$ is
\begin{equation}
- \, \frac{\mu(2\mu-3)\lambda_1}{2(\mu-1)^2(\mu-2)(4\mu-7)\Nf}
\end{equation}
which is $O(1/\Nf)$, so that in the final expression one is left with an $O(1)$
quantity involving the unknown $\lambda_1$. The other elements of the matrix
are both $O(1)$. Of course, if one naively follows this path, the correct
expression for $\lambda_1$ does not emerge, for the simple reason that a
contribution has been omitted. This arises from the delicate balancing of
powers of $\Nf$ between $s(\alpha-1,\lambda)$ and the term it multiplies in the
determinant to yield an $O(1)$ contribution. If we trace the origin of the
$O(1/\Nf)$ quantity in (3.6), it arises purely from the one loop graph of the
$A_{\mu\nu}$ equation of fig. 1. However, if one considers the higher two loop
graph of fig. 2, which would be the next term of the Dyson equation for the
photon, then it will be of the same order as (3.11), and therefore this graph
must be included. It corresponds to taking the leading term of (2.2) and the
$(x^2)^\lambda$ term of (2.3), in the replacement of lines at criticality. For
completeness, we remark that this reordering of graphs does not occur in the
pioneering work of \cite{7,8}, but does occur in other models with fermions,
such as the $O(N)$ Gross Neveu model, though the current point was not
explicitly discussed in \cite{10}.

We now amend our equations to include this extra contribution. First, we
define the value of the relevant two loop correction by
\begin{equation}
\Pi_{\mu\nu} ~+~ (x^2)^\lambda B^\prime \hat{\Pi}_{\mu\nu}
\end{equation}
where $\Pi_{\mu\nu}$ $=$ $\Pi \eta_{\mu\nu}$ $+$ $\Xi x_\mu x_\nu/x^2$ etc and
$\Pi$, $\Xi$, $\hat{\Pi}$ and $\hat{\Xi}$ denote the explicit values one
obtains when the corresponding two loop integral is computed. Thus, (3.2)
becomes
\begin{eqnarray}
0 &=& A^{-1}_{\mu\nu}(x) - \frac{4z\Nf}{(x^2)^{2\alpha-1}} \left( \eta_{\mu\nu}
- \frac{2 x_\mu x_\nu}{x^2} \right) [ 1 + 2 (x^2)^\lambda A^\prime] \nonumber
\\
&-& \frac{z^2\Nf}{(x^2)^{4\alpha+\beta-2\mu-2}} \left[ \Pi \eta_{\mu\nu}
+ \Xi \frac{x_\mu x_\nu}{x^2} \right] \nonumber \\
&-& \frac{z^2 \Nf B^\prime}{(x^2)^{4\alpha+\beta-2\mu-2-\lambda}}
\left[ \hat{\Pi} \eta_{\mu\nu} + \hat{\Xi} \frac{x_\mu x_\nu}{x^2} \right]
\end{eqnarray}
Following the procedure we discussed earlier, (3.6) is then replaced by
\begin{eqnarray}
0 &=& \frac{8(2\alpha-\lambda-2)A^\prime}{(2\alpha-\lambda-1)}
+ \left[ \frac{8n(\beta)(\alpha-1)(\mu-\beta-\lambda)(2\mu-2\beta+1)}
{(2\alpha-1)(\mu-\beta)(2\mu-2\beta-2\lambda+1)} \right. \nonumber \\
&+& \left. z \left( \hat{\Pi} + \frac{\hat{\Xi}}{2(4\alpha+\beta-2\mu
-\lambda-2)} \right) \right] B^\prime
\end{eqnarray}
where it is clear that both contributions to the $B^\prime$ term are the
same order, $O(1/\Nf)$, since $\hat{\Pi}$ $=$ $\hat{\Xi}$ $=$ $0(1)$ and
$z$ $=$ $O(1/\Nf)$.

\sect{Computation of $\hat{\Pi}_{\mu\nu}$.}

This section is devoted to evaluating $\hat{\Pi}_{\mu\nu}$ by building on our
earlier $O(1/\Nf^2)$ work, where we developed various integration rules to
treat gauge fields in the self consistency formalism of \cite{6}, as well as
to determine some two loop graphs similar to $\hat{\Pi}_{\mu\nu}$. As
mentioned, the graph of fig. 2 involves a gauge field with an exponent of
$\beta$ $-$ $\lambda$. The shift of the exponent from the original $\beta$ has
a major consequence. In the absence of such an insertion, the graph corresponds
to $\Pi_{\mu\nu}$, which was discussed in \cite{6}, and which is in fact
infinite due to divergences from vertex subgraphs. However, one can compute the
divergent piece as well as the finite piece after a suitable regulator is
introduced. In the present case, the shift to $\beta$ $-$ $\lambda$ means
$\hat{\Pi}_{\mu\nu}$ is finite and does not possess the infinities of
$\Pi_{\mu\nu}$, which will be seen from the calculation, or by realising that
the insertion alters the structure of the vertex subgraphs. If, by contrast,
one made a shift in an electron line, that graph would remain divergent due
to the divergence arising from the vertex subgraph not affected by such a
shift. Whilst $\hat{\Pi}_{\mu\nu}$ is finite and therefore does not need to
be regularized for subgraph infinites, this means its computation cannot be
carried out easily.

In other models, however, it was possible to calculate completely finite graphs
by the method of uniqueness, which was first introduced in \cite{18} and
developed in various applications later, [19-22]. The basic rule of integration
is given in fig. 3, where the Greek letter beside each line corresponds to the
critical exponent for that purely bosonic propagator. When the arbitrary
exponents, $\alpha$, $\beta$ and $\gamma$, satisfy $\alpha$ $+$ $\beta$ $+$
$\gamma$ $=$ $2\mu$, then one can perform the integration over the point of
intersection of the three external vertices to obtain the product of three
propagators with suitably adjusted exponents, and represented by a triangle in
fig. 3, where $\nu(\alpha_1, \alpha_2, \alpha_3)$ $=$ $\pi^\mu \prod_{i=1}^3
a(\alpha_i)$. For QED, it ought therefore, to be possible to calculate the
graph of fig. 2 directly using the same techniques. However, as we noted in
\cite{6}, there is no simple uniqueness rule analogous to fig. 3 for a photon
electron vertex. Indeed, for such a vertex to be unique, ie integrable, the
constraint is that the sum of the exponents is $2\mu$ $+$ $2$, but from (2.6),
$2\alpha$ $+$ $\beta$ $=$ $2\mu$ $+$ $1$ at leading order, which is one step
away from uniqueness and thus useless for our purposes. To calculate
$\Pi_{\mu\nu}$ in \cite{6}, in the absence of a suitable rule, meant rewriting
the integral first by integration by parts rules and then taking the trace over
the $\gamma$-matrices. The former step using the rules of figs. 4 and 5,
reduces the number of $\gamma$-matrices or simplifies the $A_\mu$ propagator,
ie the first term of the right side of fig. 4. (In figs. 4 and 5, the letter
$\mu$ or $\nu$ at the vertex of integration denotes the presence of a
$\gamma$-matrix there, either free or contracted with the gauge field. The
appearance of a similar letter at the end of a line corresponds to a vector
$z_\nu$ in the numerator of the integral, where $z_\mu$ is the vector joining
this external point with the integration vertex, which we represent graphically
as a long dashed line.) Having split the integral in this way, one takes the
trace and then projects out the $\eta_{\mu\nu}$ and $x_\mu x_\nu$ components so
that the graph can be written as a sum of purely bosonic two loop integrals,
where the exponents are related to those of $\Pi_{\mu\nu}$ or shifted by
integers.

We followed a similar path here to compute $\hat{\Pi}_{\mu\nu}$, and if we
denote by $I$ the contribution to $\hat{\Pi}_{\mu\nu}$ from the first term of
the right side of fig. 4 (omitting the displayed factor), and by $II$ the
remaining two vertices, we will concentrate on $\hat{\Pi}_{\mu\nu}^I$. Using
the convention $\mbox{tr} 1$ $=$ $4$, together with $\{ \gamma^\mu, \gamma^\nu
\}$ $=$ $2\eta^{\mu\nu}$ only, then
\begin{eqnarray}
\hat{\Pi}^{I \sigma}_\sigma &=& - \, 8(\mu-1) [(\mu-2) A_{3-\mu}
+ B_{2-\mu} + B_{3-\mu} - 2 C_{3-\mu} ] \\
\hat{\Pi}^I_{\sigma\rho}x^\sigma x^\rho &=& - \, 4 [ B_{2-\mu}
+(\mu-1)B_{3-\mu} + 2D_{3-\mu} + 2E_{3-\mu}]
\end{eqnarray}
where we have defined the various linear combinations of purely bosonic
graphs as
\begin{eqnarray}
A_\xi &=& 2 \la \alpha-1, \alpha, \alpha-1, \alpha, \xi \ra
- \la \alpha, \alpha, \alpha, \alpha, \xi-1 \ra \\
B_\xi &=& \la \alpha, \alpha, \alpha, \alpha, \xi \ra
+ 2 \la \alpha, \alpha-1, \alpha-1, \alpha, \xi \ra \nonumber \\
&-& 4 \la \alpha-1, \alpha, \alpha, \alpha, \xi \ra
+ \la \alpha, \alpha, \alpha, \alpha, \xi-1 \ra \\
C_\xi &=& \la \alpha-1, \alpha, \alpha, \alpha-1, \xi-1 \ra
- \la \alpha, \alpha, \alpha-1, \alpha-1, \xi \ra \\
D_\xi &=& 2 \la \alpha-1, \alpha, \alpha-2, \alpha, \xi \ra
- \la \alpha-1, \alpha, \alpha-1, \alpha, \xi-1 \ra \nonumber \\
&-& \la \alpha-1, \alpha, \alpha-1, \alpha, \xi \ra \\
E_\xi &=& \la \alpha-1, \alpha-1, \alpha, \alpha, \xi \ra
- 2 \la \alpha-1, \alpha-1, \alpha-1, \alpha, \xi \ra
\end{eqnarray}
where the general two loop integral $\la \alpha_1, \alpha_2, \alpha_3,
\alpha_4, \alpha_5 \ra$ is defined in fig. 6. The reason for these particular
linear combinations will become apparent later.

Whilst we pointed out that $\hat{\Pi}_{\mu\nu}$ itself was finite, with respect
to vertex subgraph divergences, the various integrals making up the definitions
in (4.3)-(4.7) are in fact divergent when one substitutes the value $\alpha$
$=$ $\mu$, which is the leading order value, with respect to $\Nf$, we are
interested in. The reason for this  divergence is that $\mu$ is the
anti-uniqueness value of a bosonic field, \cite{8}, and in, say (2.4), we would
have terms such as $a(\mu+\eta/2)$ $\sim$ $- \, 2/(\eta\Gamma(\mu))$. The main
point to realise is that whilst individual graphs such as $\la \alpha, \alpha,
\alpha, \alpha, 3-\mu \ra$, for example, suffer from this infinity, in their
sum the infinities will cancel and it is therefore important to arrange the
calculation in such a way that this happens. Indeed this is the reason for the
choice of combinations in (4.3)-(4.7).

We now illustrate this but note that first we need to rewite each integral in
such a way that the divergence, $1/(\mu-\alpha)$, is displayed explicitly. One
clue to this is to realise that if each integral could be written in terms of
integrals with no anti-uniqueness then that graph would itself be finite, with,
hopefully, a coefficient involving $1/(\mu-\alpha)$. Recently, various
recursion relations to reduce the exponents of lines in two loop graphs of the
form of fig. 6 have been given in \cite{22} and used in \cite{6}. As they as
completely general, they can be applied to the present problem and we have
illustrated those we used in figs. 7-10, where the $+$ or $-$ indicates the
addition or subtraction of $1$ to the exponent of the corresponding propagator
of the left side. Also, a propagator joining the endpoints corresponds to
multiplying the integral by appropriate powers of $x^2$, to ensure the
dimensions of each term is consistent.

To illustrate these points, we derive an expression for $C_\xi$ by using figs.
7 and 8, respectively, for each term of (4.5), to find
\begin{eqnarray}
C_\xi &=& \frac{(4\alpha+\xi-2\mu-4)(3\mu-4\alpha-\xi+3)}{(\alpha-1)^2}
\nonumber \\
&& \times \la \alpha-1, \alpha-1, \alpha-1, \alpha-1, \xi \ra \nonumber \\
&+& \frac{\xi(\xi+1-\mu)}{(\alpha-1)^2} \la \alpha-1, \alpha-1, \alpha-1,
\alpha-1, \xi+1 \ra
\end{eqnarray}
where the two terms involving the divergent integral $\la \alpha-1,
\alpha-1, \alpha-1, \alpha, \xi \ra$ have cancelled. From (4.8), we see
that $C_\xi$ involves graphs with no anti-uniqueness when $\alpha$ $=$
$\mu$ and it is therefore finite. For $C_{3-\mu}$, only the second term of
(4.8) survives and we are left to evaluate $\la \mu-1, \mu-1, \mu-1, \mu-1,
4-\mu \ra$ which can be achieved by making the successive transformations
$\nearrow$ and $\searrow$ in the notation of \cite{8}, to yield
$\la \mu-2, 2, 2, \mu-1, \mu-2 \ra$. Applying the rule of fig. 9 to this
leaves the graph $\la \mu-1, 1, 2, \mu-1, \mu-1 \ra$, which has a unique
vertex. Thus,
\begin{equation}
C_{3-\mu} ~=~ \frac{2(\mu-3)\pi^{2\mu}}{(\mu-2)\Gamma^2(\mu)}
\end{equation}
In effect the process to evaluate the combinations is two stage. The first
involves rewriting the constituent integrals to yield finite ones, before
evaluating for the specific cases of $\xi$ $=$ $2$ $-$ $\mu$ or $3$ $-$
$\mu$, we require. In the appendix, we have listed the values of the basic
finite integrals analogous to $\la \mu-1, \mu-1, \mu-1, \mu-1, 4-\mu \ra$.
They were also deduced by the recursion relations, which allows one to
rewite an unknown integral in terms of others in the sequence which have
already been evaluated. Further, we have also listed the expressions for
the remaining combinations, (4.3)-(4.7), though we note their derivation
was more involved than $C_\xi$.

One graph, $\la  \alpha, \alpha-1, \alpha-1, \alpha-1, \xi \ra$, which occurs
in some of the combinations but does not contribute to $C_\xi$, deserves
special mention. Whilst it is divergent, we had to use the following circuitous
argument to write it with explicit $1/(\mu-\alpha)$ factors. First, applying
the rule of fig. 9, gives
\begin{equation}
\la \alpha-1, \alpha, \alpha-2, \alpha-1, \xi+1 \ra
+ \frac{(\mu-\alpha-\xi)}{(\alpha-1)} \la \alpha-1, \alpha-1, \alpha-1,
\alpha-1, \xi+1 \ra
\end{equation}
Next, we apply the transformation of $\leftarrow$ of \cite{8} to $\la
\alpha-1, \alpha-2, \alpha, \alpha-1, \xi+1 \ra$, use a recursion relation
before applying the inverse, $\leftarrow$. Of the three integrals which now
result, two are finite, $\la \alpha-1, \alpha-2, \alpha-1, \alpha-1, \xi+1
\ra$ and $\la \alpha-1, \alpha-3, \alpha-1, \alpha-1, \xi+3 \ra$ with a
factor $1/(\mu-\alpha)$ in their coefficient, whilst the remaining,
$\la \alpha, \alpha-1, \alpha-2, \alpha-1, \xi+1 \ra$ still has a hidden
divergence. Applying the rule of fig. 10 to this latter integral, the
integral we are interested in emerges, whilst the remainder are finite. Thus,
collecting the various contributions and rearranging, we have
\begin{eqnarray}
&& b(\alpha,\xi) \la \alpha-1, \alpha-1, \alpha-1, \alpha, \xi \ra \nonumber \\
&&~=~ c(\alpha,\xi) \la \alpha-1, \alpha-1, \alpha-1, \alpha-1, \xi+1 \ra
\nonumber \\
&&~+~ \frac{(\xi+1)d(\alpha,\xi)}{(\mu-\alpha)} \la \alpha-1, \alpha-3,
\alpha-1, \alpha-1, \xi+2 \ra \nonumber \\
&&~+~ \frac{(2\alpha-\mu-3)d(\alpha,\xi)}{(2\alpha+\xi-\mu-1)(2\mu-2\alpha
-\xi)} \la \alpha-1, \alpha-1, \alpha-2, \alpha-1, \xi+1 \ra \nonumber \\
&&~+~ \frac{(\xi+1)(2\alpha-\mu-3)}{(\mu-\alpha)(\alpha-1)}
\la \alpha-1, \alpha-2, \alpha-1, \alpha-1, \xi+2 \ra
\end{eqnarray}
where
\begin{eqnarray}
b(\alpha,\xi) &=& 1 - \frac{(2\alpha-\mu-3)(\alpha-2)}{(2\alpha+\xi-\mu-2)
(2\alpha+\xi-\mu-1)(2\mu-2\alpha-\xi)} \nonumber \\
c(\alpha,\xi) &=& \frac{(\mu-\alpha-\xi)}{(\alpha-1)}
+ \frac{(2\alpha-\mu-3)(\alpha-2)(2\mu-2\alpha-\xi+1)}{(\alpha-1)
(2\alpha+\xi-\mu-1)(2\mu-2\alpha-\xi)} \nonumber \\
d(\alpha,\xi) &=& \frac{(3\mu-4\alpha-\xi+3)(4\alpha-2\mu+\xi-4)}
{(\alpha-1)(2\alpha+\xi-\mu-2)}
\end{eqnarray}
and we  have displayed the $1/(\mu-\alpha)$ pieces explicitly. Thus, simply
\begin{eqnarray}
[(\mu-\alpha)\la \alpha-1, \alpha-1, \alpha-1, \alpha, 3-\mu \ra ]
|_{\alpha\,=\,\mu} &=& \frac{(\mu-1)\pi^{2\mu}}{(\mu-2)\Gamma^2(\mu)} \\
\left[(\mu-\alpha)\la \alpha-1, \alpha-1, \alpha-1, \alpha, 4-\mu \ra \right]
|_{\alpha\,=\,\mu} &=& \frac{(\mu-1)\pi^{2\mu}}{2(\mu-3)\Gamma^2(\mu)} ~~
\end{eqnarray}
which completes the basic results required to compute (4.3)-(4.7). Thus from
the results listed in the appendix,
\begin{eqnarray}
\hat{\Pi}_\sigma^{I \sigma} &=& - \, \frac{8\pi^{2\mu}(5\mu^2-11\mu+4)}
{(\mu-1)(\mu-2)\Gamma^2(\mu)} \\
\hat{\Pi}^I_{\sigma \rho} x^\sigma x^\rho &=& - \, \frac{4\pi^{2\mu}
(2\mu^3-13\mu^2+21\mu-8)}{(\mu-1)(\mu-2)\Gamma^2(\mu)}
\end{eqnarray}

To complete the calculation, the contribution from $\hat{\Pi}^{II}_{\mu\nu}$ is
simplified considerably by noticing that from the integration by parts rules of
figs. 4 and 5, the factor multiplying $\hat{\Pi}^{II}_{\mu\nu}$ is
$(\mu-\alpha)^2$, so that in applying the recursion relations on its
constituent bosonic integrals, very few will survive in the limit as $\alpha$
$\rightarrow$ $\mu$. Essentially, one need only isolate those terms which lead
to $(\mu-\alpha) \la \alpha-1, \alpha-1, \alpha-1, \alpha, \xi \ra$. We found,
\begin{equation}
\hat{\Pi}^{II}_{\mu\nu} ~=~ \frac{16\pi^{2\mu}}{(\mu-2)\Gamma^2(\mu)}
\left[ \eta_{\mu\nu} - \frac{2x_\mu x_\nu}{x^2} \right]
\end{equation}
Thus, overall,
\begin{eqnarray}
\hat{\Pi} &=& \frac{8\pi^{2\mu}(2\mu^2-21\mu+25)}{(4\mu-7)(\mu-1)\Gamma^2
(\mu)} \\
\hat{\Xi} &=& - \, 2 \hat{\Pi} - \frac{16\pi^{2\mu}(2\mu-3)(\mu-2)(\mu-3)}
{(4\mu-7)(\mu-1)\Gamma^2(\mu)}
\end{eqnarray}
One final point to note is that in the list of finite integrals we used,
given in the appendix, the function $ChT(1,1)$ $=$
$3\pi^{2\mu}a(2\mu-2)$$a(1)$$[\psi^\prime(\mu-1)-\psi^\prime(1)]$ appears,
where $ChT(\alpha,\beta)$ $=$ $\la \alpha, \mu-1, \mu-1, \beta, \mu-1 \ra$ in
the notation of \cite{8} and $\psi(x)$ is the logarithmic derivative of the
$\Gamma$-function. It turns out that this function cancels in the final
expression for $\hat{\Pi}_{\mu\nu}$.

\sect{Discussion.}

Having calculated the graph of fig. 2, we can now evaluate the determinant of
the matrix formed by (3.5) and (3.11). With the leading values $\alpha$ $=$
$\mu$, $\beta$ $=$ $1$ and $\lambda$ $=$ $\mu$ $-$ $2$ for the terms not
involving $\hat{\Pi}_{\mu\nu}$ and noting, \cite{6},
\begin{equation}
z ~=~ - \, \frac{\mu(2\mu-3)\Gamma^2(\mu) \eta_1}{4\pi^{2\mu}
(2\mu-1)(\mu-2) \Nf} + O \left( \frac{1}{\Nf^2} \right)
\end{equation}
a little algebra gives
\begin{equation}
\lambda ~=~ \mu - 2 - \frac{(2\mu-3)(\mu-3)\eta_1}{\Nf} + O \left(
\frac{1}{\Nf^2} \right)
\end{equation}
We make several comments. First, it is easy to see that in three dimensions the
$O(1/\Nf)$ correction vanishes, so that $\beta^\prime(g_c)$ $=$ $1$ $+$
$O(1/\Nf^2)$. Second, one question which arises in applying the formalism of
\cite{7,8} concerns which perturbative renormalization scheme the critical
exponents relate to. In the conventional approach to calculating $\beta(g)$,
the first few terms of the perturbative series are independent of the
renormalization scheme one chooses. Beyond this, at three and higher loops, one
must specify the scheme, ie the manner in which one removes the divergences to
obtain the finte Green's functions of the theory. One such scheme is minimal
subtraction, where only the divergent terms and no finite pieces are absorbed
into renormalization constants. Moreover, this scheme is a mass independent
scheme. In other models, as well as QED, the exponents one calculates by the
method we used here, can be related to the corresponding renormalization group
functions and therefore compared with these functions to the orders they are
known. So far they are in agreement with $\overline{\mbox{MS}}$ scheme
functions, but there remains the question of whether this is true to all orders
in perturbation theory. Intuitively, this is perhaps obvious because the
exponents are deduced at a point of the theory where the mass vanishes, so that
they would have to correspond to a mass independent scheme. As the
$\beta$-function for QED has been explicitly calculated in the large $\Nf$
expansion in an $\mbox{MS}$ scheme, \cite{3,4}, we can thus compare the
exponent we obtained, to that derived from the renormalization of \cite{3,4},
and find both are in agreement.  By the same reasoning, we assume that similar
exponents in other models calculated using the method of \cite{7,8} give
$\overline{\mbox{MS}}$ information to all orders in perturbation theory.

The remainder of our discussion is devoted to the algorithm to go beyond (5.2),
based on the self consistency approach given here. First, the terms of (3.5)
and (3.11) have to be expanded to the next order in $1/\Nf$. For the one loop
graphs and $2$-point scaling functions, this is straightforward and, moreover,
$z$ has been computed to $O(1/\Nf^2)$ in \cite{6}. Next, the correction to
$\hat{\Pi}_{\mu\nu}$ has to be calculated, but this is now possible given our
discussion of sect. 4 where the calculation of the linear combinations,
(4.3)-(4.7), were discussed in general terms and, therefore, much of the
machinery is already in place to treat the corrections. The main computing
effort lies, though, in obtaining the values of the higher order graphs of
figs. 11 and 12 and others whose evaluation lie beyond the scope of the
present paper. These graphs fall into two classes. For the two loop graph, we
must consider corrections to both the electron and photon propagators, (2.2)
and (2.3). As we remarked earlier, an insertion on the completely internal line
yields a finite graph, whilst that on a line joining an external vertex will be
divergent. However, the procedure to regularize and renormalize these
divergences was given in \cite{6}. For the graphs of fig. 12, which give
corrections by the same argument that $\hat{\Pi}_{\mu\nu}$ was needed for
$\lambda_1$, one need only consider corrections to the photon propagator, which
means the four loop graph is finite with respect to vertex subgraph
divergences.

We conclude with several remarks. First, we believe that it is now viable to
push calculations in QED to $O(1/\Nf^2)$ in arbitrary dimensions. We have
stated earlier why such calculations are required for four dimensional
perturbation theory, but note that, of course, we will simultaneously deduce
useful information about the three dimensional model. Second, as the method
will be very involved, it might be more appropriate to examine the
practicalities in a model with similar properties, but with simpler
$O(1/\Nf^2)$ structure, such as the $O(N)$ Gross Neveu model, where $\lambda_2$
is unknown too. Also, the basic integrals which occur in that model will be
relevant for the QED calculation.
\newpage
\appendix
\sect{Various finite integrals.}
In this appendix, we summarize the various building blocks required for
$\hat{\Pi}_{\mu\nu}$, giving the exact values of various finite integrals
first.
\begin{eqnarray}
&&\la \mu-1, \mu-1, \mu-1, \mu-1, 4-\mu \ra ~=~ \frac{\pi^{2\mu}(\mu-1)^2}
{(\mu-2)^2\Gamma^2(\mu)} \\
&&\la \mu-2, \mu-1, \mu-1, \mu-1, 4-\mu \ra \nonumber \\
&& ~~=~ - \, \frac{a(4-\mu)}{a^3(1)} \left[ \frac{\pi^{2\mu}\Gamma(\mu-1)
\Gamma(2-\mu)}{\Gamma(2\mu-3)} + (\mu-2)^2 ChT(1,1) \right] \\
&&\la \mu-1, \mu-1, \mu-1, \mu-1, 5-\mu \ra ~=~ \frac{\pi^{2\mu} (\mu-1)^3}
{2(\mu-3)(\mu-2)^2 \Gamma^2(\mu)} \\
&& \la \mu-2, \mu-1, \mu-1, \mu-1, 5-\mu \ra ~=~ \frac{\pi^{2\mu} (\mu-1)^2}
{2(\mu-2)(\mu-3)\Gamma^2(\mu)} \\
&&\la \mu-3, \mu-1, \mu-1, \mu-1, 5-\mu \ra \nonumber \\
&& ~~=~ \frac{a(5-\mu)}{a(1)a^2(2)} \left[ (\mu-3)^2 ChT(1,1)
+ \frac{\pi^{2\mu}(\mu^2-7\mu+11)a(1)}{(\mu-2)^3a(3-\mu)}
\right] \\
&&\la \mu-2, \mu-1, \mu-1, \mu-2, 5-\mu \ra \nonumber \\
&& ~~=~ \frac{a(5-\mu)(\mu-2)}{a^3(2)} \left[ ChT(1,1) - \frac{\pi^{2\mu}
(\mu-3)a(1)}{(\mu-2)^3a(3-\mu)} \right] \\
&& \la \mu-2, \mu-1, \mu-2, \mu-1, 5-\mu \ra \nonumber \\
&& ~~=~ \frac{a(5-\mu)}{a^3(2)} \left[ ChT(1,1)
+ \frac{\pi^{2\mu}(2\mu^2 - 12\mu + 17)a(2)}{(\mu-2)^2a(3-\mu)} \right] \\
&& \la \mu-1, \mu-1, \mu-1, \mu-1, 6-\mu \ra \nonumber \\
&& ~~~~~~~~~~~~~~~~~~~=~~ \frac{\pi^{2\mu} (\mu-1)^2 (5\mu^3 - 27 \mu^2
+ 40 \mu -24)}{18 (\mu-2)^2(\mu-3)(\mu-4)\Gamma^2(\mu)} \\
&& \la \mu-2, \mu-1, \mu-1, \mu-1, 6-\mu \ra \nonumber \\
&& ~~~~~~~~~~~~~~~~~~~=~~ \frac{\pi^{2\mu} (\mu-1)^2 (3\mu^2 - 14 \mu + 12)}
{12(\mu-2)(\mu-3)^2(\mu-4) \Gamma^2(\mu)} \\
&& \la \mu-2, \mu-1, \mu-2, \mu-1, 6-\mu \ra ~=~ \! \frac{\pi^{2\mu}(\mu-4)
(\mu-1)^2}{4(\mu-2)(\mu-3)^2 \Gamma^2(\mu)} ~~~~ \\
&& \la \mu-2, \mu-1, \mu-1, \mu-2, 6-\mu \ra ~=~ \! \frac{\pi^{2\mu}(\mu-1)^2}
{6(\mu-3)(\mu-4) \Gamma^2(\mu)} \\
&& \la \mu-3, \mu-1, \mu-1, \mu-1, 6-\mu \ra ~=~ \! \frac{\pi^{2\mu} (\mu-1)^2}
{3(\mu-2)(\mu-4) \Gamma^2(\mu)} ~~~~
\end{eqnarray}
Next, we give the explicit expressions for each finite linear combination of
divergent graphs defined in sect. 4, for arbitrary $\xi$.
\begin{eqnarray}
A_\xi \!\! &=& \!\! \frac{2(\mu-\alpha)(2\mu-2\alpha-\xi)(2\alpha+\xi-\mu-1)}
{(\alpha -1)^3} \la \alpha-1, \alpha-1, \alpha, \alpha-1, \xi+1 \ra \nonumber
\\
&+& [(2\mu-\xi-\alpha-1)(\mu-\alpha-\xi)-(\mu-\alpha)(\alpha-1)](3\mu-4\alpha
-\xi+2) \nonumber \\
&& \times ~ \frac{(4\alpha+\xi-2\mu-3)}{(\alpha-1)^4}
\la \alpha-1, \alpha-1, \alpha-1, \alpha-1, \xi+1 \ra
\end{eqnarray}
\begin{eqnarray}
B_\xi &=& \frac{(4\alpha+\xi-2\mu-3)(3\mu-4\alpha-\xi+2)}{(\alpha-1)^4}
\left[ \frac{(\mu-1)\xi(\xi-\mu+1)}{(2\alpha+\xi-\mu-1)} \right. \nonumber \\
&& +~ \left. 2(\mu-\alpha)^2 \frac{}{} \right] \la \alpha-1, \alpha-1,
\alpha-1, \alpha-1, \xi+1 \ra \nonumber \\
&+& \frac{2(\mu-\alpha)}{(\alpha-1)^3} \left[ 2(\mu-\alpha)(2\alpha-\mu
-1) - \frac{(\mu-1)\xi(\xi-\mu+1)}{(2\alpha+\xi-\mu-1)} \right] \nonumber \\
&&\times ~\la \alpha-1, \alpha-1, \alpha, \alpha-1, \xi+1 \ra
\end{eqnarray}
\begin{eqnarray}
C_\xi &=& \frac{(4\alpha+\xi-2\mu-4)(3\mu-4\alpha-\xi+3)}{(\alpha-1)^2}
\la \alpha-1, \alpha-1, \alpha-1, \alpha-1, \xi \ra \nonumber \\
&-& \frac{\xi(\mu-\xi-1)}{(\alpha-1)^2} \la \alpha-1, \alpha-1, \alpha-1,
\alpha-1, \xi+1 \ra
\end{eqnarray}
\begin{eqnarray}
D_\xi &=& - \, \frac{2(\mu-\alpha)(\alpha+\xi-2)}{(\alpha-1)(\alpha-2)}
\la \alpha-1, \alpha-1, \alpha, \alpha-1, \xi \ra \nonumber \\
&+& \frac{\xi}{(\alpha-1)} \left[ 2 + \frac{(2\alpha+\xi-\mu-3)
(\mu-\alpha-\xi)}{(\alpha-1)(\alpha-2)} \right] \nonumber \\
&&\times~ \la \alpha-1, \alpha-1, \alpha-1, \alpha-1, \xi+1 \ra \nonumber \\
&+& \frac{2(\mu-\alpha)(2\mu-2\alpha-\xi+2)(2\alpha+\xi-\mu-3)}
{(\alpha-1)(\alpha-2)(2\alpha+\xi-\mu-2)(2\mu-2\alpha-\xi+1)} \nonumber \\
&&\times~ \la \alpha-1, \alpha-1, \alpha, \alpha-1, \xi-1 \ra \nonumber \\
&+& f(\alpha, \xi) \, \la \alpha-1, \alpha-1, \alpha-1, \alpha-1,
\xi \ra
\end{eqnarray}
\begin{eqnarray}
E_\xi &=& \frac{\xi(\mu-\xi-1)}{(\alpha-1)^2} \la \alpha-1, \alpha-1,
\alpha-1, \alpha-1, \xi+1 \ra \nonumber \\
&-& \frac{2(\mu-\alpha)}{(\alpha-1)} \la \alpha-1, \alpha-1, \alpha,
\alpha-1, \xi \ra
\end{eqnarray}
where we have set
\begin{eqnarray}
f(\alpha, \xi) \! &=& \! \! \frac{(3\mu-4\alpha-\xi+3)(\mu-\alpha-\xi+1)
(2\mu-2\alpha-\xi+2)}{(\alpha-1)^2(\alpha-2)(2\mu-2\alpha-\xi+1)
(2\alpha+\xi-\mu-2)} \nonumber \\
&\times& \! \! [ \frac{}{} (\alpha-1)(2\mu-2\alpha-\xi+1)
(4\alpha+2\xi-3\mu-4)(2\alpha+\xi-4) ~~~~~~ \nonumber \\
&+& (\mu-\alpha-\xi+1)(2\mu-2\alpha-\xi+2) \nonumber \\
&&\!\!\!\!\!\! \times \,\, [(\mu-\xi)(2\alpha+\xi-\mu-3)
- (4\alpha+2\xi-3\mu-4)(\mu-1)]]
\end{eqnarray}
Finally, the finite expressions for the various combinations we needed are,
\begin{eqnarray}
A_{2-\mu} &=& \frac{2\pi^{2\mu}}{(\mu-1)^2 \Gamma^2(\mu)} \\
A_{3-\mu} &=& - \, \frac{2\pi^{2\mu}(\mu^2 - 4\mu + 2)}{(\mu-1)^2(\mu-2)
\Gamma^2(\mu)} \\
B_{2-\mu} &=& \frac{2\pi^{2\mu}(2\mu-3)}{(\mu-1)\Gamma^2(\mu)} \\
B_{3-\mu} &=& \frac{\pi^{2\mu}(2\mu^2-7\mu+4)}{(\mu-1)(\mu-2)\Gamma^2(\mu)} \\
C_{3-\mu} &=& \frac{2\pi^{2\mu}(\mu-3)}{(\mu-2)\Gamma^2(\mu)} \\
D_{3-\mu} &=& E_{3-\mu} ~=~ - \, \frac{2\pi^{2\mu}}{\Gamma^2(\mu)}
\end{eqnarray}

\newpage
\noindent
{\Large {\bf Figure Captions.}}
\begin{description}
\item[Fig. 1.] Skeleton Dyson equations for $\psi^i$ and $A_\mu$.
\item[Fig. 2.] Additional graph contributing to $\lambda_1$.
\item[Fig. 3.] Uniqueness rule for a bosonic vertex.
\item[Fig. 4.] Integration by parts rule for a gauge vertex.
\item[Fig. 5.] Additional integration by parts rule.
\item[Fig. 6.] General two loop self energy graph.
\item[Fig. 7.] Recursion relation for two loop self energy which increases
$t_1$ by $1$.
\item[Fig. 8.] Recursion relation for two loop self energy which reduces
$s_1$ by $1$.
\item[Fig. 9.] Recursion relation for two loop self energy which increases
$t_2$ by $1$.
\item[Fig. 10.] Recursion relation for two loop self energy which reduces
$t_1$ by $1$.
\item[Fig. 11.] Higher order corrections for electron self energy.
\item[Fig. 12.] Higher order corrections for photon self energy.
\end{description}
\end{document}